\documentclass[]{spie}  

\usepackage{amsmath,amsfonts,amssymb}
\usepackage{graphicx}
\usepackage[colorlinks=true, allcolors=blue]{hyperref}

\usepackage[normalem]{ulem}
\usepackage{color}

\newcommand{\um}{$\mu {\rm m}$}

\title{Optical Characterization \& Testbed Development for $\mu$-Spec Integrated Spectrometers}

\author[a]{Maryam Rahmani}
\author[a]{Alyssa Barlis}
\author[a]{Emily M. Barrentine}
\author[a]{Ari D. Brown}
\author[a]{Berhanu T. Bulcha}
\author[a]{Giuseppe Cataldo}
\author[a,d]{Jake Connors}
\author[a]{Negar Ehsan}
\author[a]{Thomas M. Essinger-Hileman}
\author[a,c]{Henry Grant}
\author[a]{James Hays-Wehle}
\author[a]{Wen-Ting Hsieh}
\author[a]{Vilem Mikula}
\author[a]{S. Harvey Moseley}
\author[a]{Omid Noroozian}
\author[b]{Trevor R. Oxholm}
\author[a]{Manuel A. Quijada}
\author[a]{Jessica Patel}
\author[a]{Thomas R. Stevenson}
\author[a]{Eric R. Switzer}
\author[e]{Carole Tucker}
\author[a]{Kongpop U-Yen}
\author[a,c]{Carolyn Volpert}
\author[a]{Edward J. Wollack}
\affil[a]{NASA Goddard Space Flight Center, Greenbelt, MD, USA}
\affil[b]{University of Wisconsin-Madison, Madison, WI, USA}
\affil[c]{University of Maryland, College Park, MD, USA}
\affil[d]{National Institute of Standards and Technology, Boulder, CO, USA}
\affil[e]{Cardiff University, Cardiff, Wales, UK}

\authorinfo{Further author information: (Send correspondence to M.R.)\\M.R.: E-mail: maryam.rahmani@nasa.gov}

\begin{document} 
\maketitle
\begin{abstract}

This paper describes a cryogenic optical testbed developed to characterize $\mu$-Spec spectrometers in a dedicated dilution refrigerator (DR) system. $\mu$-Spec is a far-infrared integrated spectrometer that is an analog to a Rowland-type grating spectrometer. It employs a single-crystal silicon substrate with niobium microstrip lines and aluminum kinetic inductance detectors (KIDs). Current designs with a resolution of $R=\lambda/\Delta \lambda=512$ are in fabrication for the EXCLAIM (Experiment for Cryogenic Large Aperture Intensity Mapping) balloon mission. The primary spectrometer performance and design parameters are efficiency, NEP, inter-channel isolation, spectral resolution, and frequency response for each channel. Here we present the development and design of an optical characterization facility and preliminary validation of that facility with earlier prototype R=64 devices. We have conducted and describe initial optical measurements of $R=64$ devices using a swept photomixer line source. We also discuss the test plan for optical characterization of the EXCLAIM $R=512$ $\mu$-Spec devices in this new testbed.

\end{abstract}

\keywords{Integrated spectrometers, far-infrared, KIDs, optical characterization, photomixer}

\section{INTRODUCTION}
\label{sec:intro}  
One of the primary areas of research in far-infrared astrophysics is the measurement of molecular emission lines to understand galactic, star, and planetary formation. The observation of these fine-structure lines of the abundant elements C, N, and O provides an opportunity to trace obscured star formation and active galactic nuclei (AGN) activity to the high redshift universe. Carbon monoxide and singly-ionized carbon lines trace star formation, promising to shed light on why the star formation rate has declined by a factor of 20 since redshift $z=3.5$, while the dark matter has continued to cluster \cite{switzer2021experiment}. Other emission lines that trace star and planetary system formation are [OIII] and [NIII], whose ratios trace the rise of metals from re-ionization to $z=1$. [CII] and OH also help characterize the trade-off between supermassive black holes (SMBH) and star formation. Further, the $538\,\mu {\rm m}$ water line provides information on the water mass in all evolutionary stages of protoplanetary systems. All of this rich science calls for the development of highly sensitive and capable far-infrared spectrometers.

\begin{figure} [ht]
  \begin{center}
  \begin{tabular}{c} 
  \includegraphics[height=4.5cm]{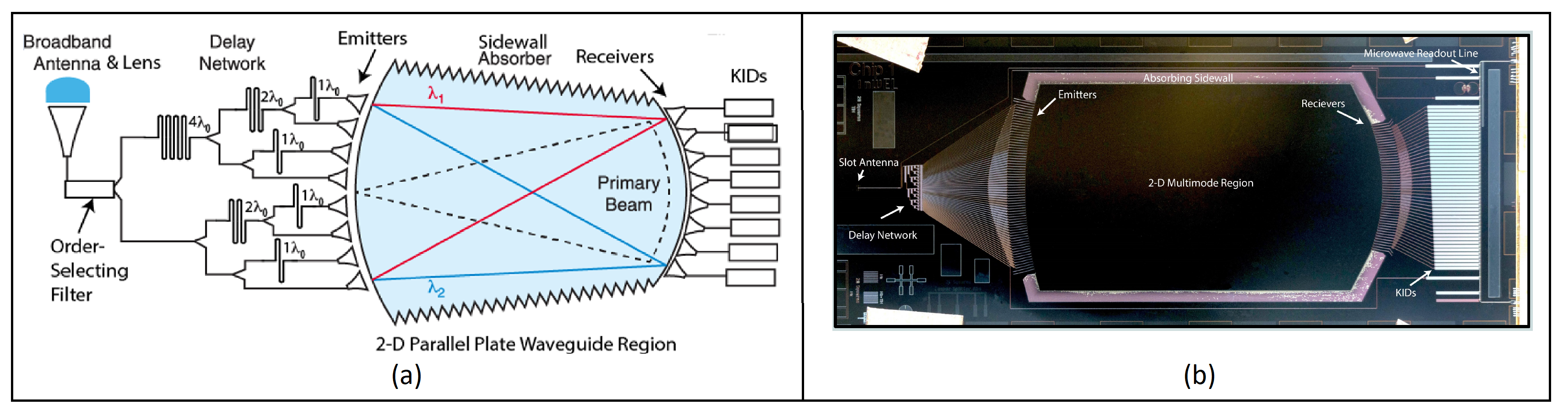}
  \end{tabular}
  \end{center}
  \caption[example] 
  { \label{fig:u_spec_block_diagram_h} 
a) A cartoon of $\mu$-Spec indicating its different components, b) image  of a fabricated R=64 $\mu$-Spec prototype. }
  \end{figure}

$\mu$-Spec is an analog to a far-infrared diffraction grating spectrometer implemented on a silicon chip, which reduces the size of the spectrometer by an order of magnitude. Similar to other integrated on-chip far-infrared spectrometers such as DESHIMA \cite{endo2012development}, SuperSpec \cite{karkare2020full}, FBS/SPT-SLIM \cite{robson2022simulation}, CAMELS \cite{thomas2014cambridge}, and SOFTS \cite{basu2020superconducting}, $\mu$-Spec\cite{cataldo2019second,cataldo2020overview,mirzaei2020mu} is a cryogenic spectrometer that utilizes superconducting planar transmission lines to transmit and spectrally sort the optical signal, superconducting microwave kinetic inductance detectors (KIDs) for detection, and operates at $\sim$100 mK temperatures. It incorporates a single-crystal silicon dielectric and superconducting niobium planar transmission lines. A cartoon figure of the $\mu$-Spec approach is shown in Figure~ \ref{fig:u_spec_block_diagram_h}. Far-infrared light from the telescope (or laboratory testbed) optics illuminates an anti-reflection (AR)-coated hyper-hemispherical Si lenslet that is coupled to a broadband dipole slot antenna. This beam is transferred to a microstrip delay network consisting of a binary tree of microstrip transmission line meanders, which synthesizes the diffraction grating and produces a linear phase delay. Then the wave is launched through emitting feeds into a 2D parallel-plate waveguide region acting as a spatial beam combiner. At the opposing side of the focal plane, receiver feeds Nyquist-sample the spectral response and couple the signal into the KIDs. Reflections and stray light incident along the sidewall of the parallel-plate waveguide region are terminated by an absorber structure \cite{bulcha2018electromagnetic}. KIDs are superconducting detectors patterned into planar microwave resonator circuits and can achieve near photon-background limited sensitivities \cite{baselmans2017kilo,hailey2021kinetic}.
KIDs respond to light with a shift in resonance frequency based on a kinetic inductance effect \cite{mazin2009microwave,zmuidzinas2012superconducting}. Every KID is designed and tuned to a unique microwave resonance frequency and is read out by a comb of microwave probe tones through a microwave feedline. A single microwave readout line is required for each spectrometer, thanks to continuous advances in microwave multiplexing readout technology \cite{bradley2021advancements,meixner2019origins,sinclair2020development}.

$\mu$-Specs have the performance characteristics needed for future balloon and space far-infrared spectroscopy missions. They can reach high efficiency and moderate resolution (up to R $\sim$1500 \cite{uSpec_Barrentine_2016}) by using low-loss superconducting transmission lines on single-crystal Si dielectric \cite{noroozian2015mu,barrentine2015overview}. Most importantly, their compact design enables a reduction in instrument size, weight and mass. The Experiment for Cryogenic Large Aperture Intensity Mapping (EXCLAIM), features a second-generation $\mu$-Spec design customized for a $555$-$714$\,\um\, band, a resolving power R=512, and the optical loading conditions at balloon float altitude \cite{cataldo2019second} with an expected NEP of $8 \times 10^{-19} {\rm W}/\sqrt{\rm Hz}$. The EXCLAIM $\mu$-Spec design is based on a first-generation prototype R=64 $\mu$-Spec design operating from $500$-$750$\,\um\, with lower sensitivity\cite{noroozian2015mu,cataldo2014micro}. EXCLAIM is a balloon-borne telescope and is designed to observe star formation by observing several emission lines from CO and [CII] \cite{EXCLAIM_Switzer_JATIS_2021, switzer2021experiment}.

Future optical and dark testing of $\mu$-Spec spectrometers will be carried out in a dilution refrigerator testbed with two new custom optical test setups. This paper presents the primary approaches and design of these new facilities and their initial validation. Section \ref{sec:DRTestbed} presents an overview of the optical test facility and its capabilities. In section \ref{sec:externalOpticalTestbed}, more details of one of the principal facilities that allow for characterization of the spectrometers with an external photomixer source are provided. In section \ref{sec:TestplanR64} the test plan for measuring a variety of the spectrometer parameters, including resolution, stray light, crosstalk, and time constant are briefly discussed. In section \ref{sec:Results}, preliminary results validating the testbed with the $R=64$ prototype devices are presented.

\section{DR Optical Testbed Overview} 
\label{sec:DRTestbed}
\begin{figure} [ht]
  \begin{center}
  \begin{tabular}{c} 
  \includegraphics[height=10cm]{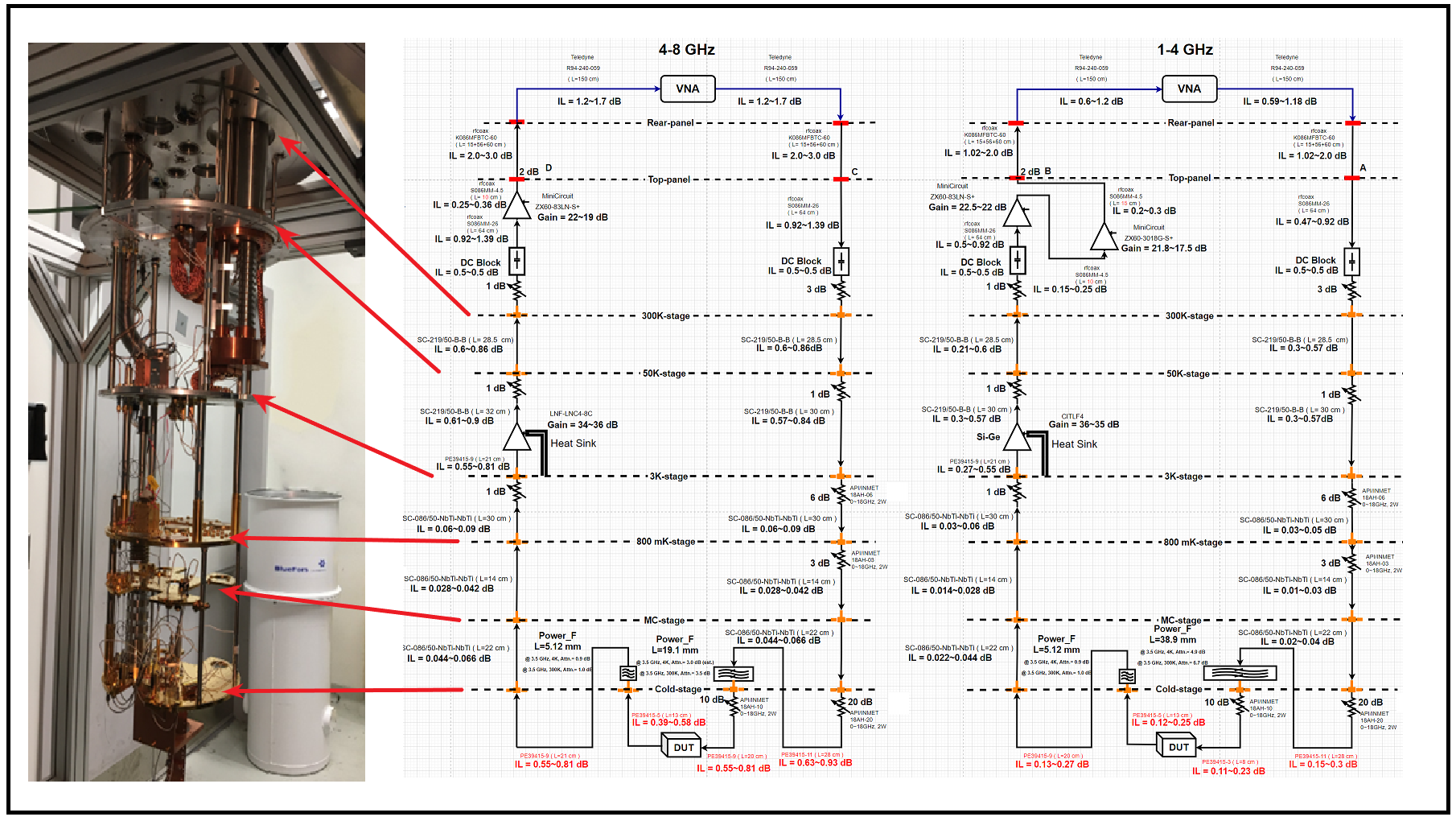}
  \end{tabular}
  \end{center}
  \caption[example] 
  { \label{fig:RF_diagram} 
Diagram of the two RF channels in the DR testbed and picture of the open cryostat stages showing where the RF lines are routed through the temperature stages.}
  \end{figure}

 The DR system employed is a Blue Fors cryogen-free (BF-LD250) dilution refrigerator, shown in Figure \ref{fig:RF_diagram}. The system provides continuous cooling power at temperatures down to $\sim$10 mK. As shown in Figure~\ref{fig:RF_diagram}-a, the spectrometers are mounted at the $\sim10$-$100$ mK temperature mixing chamber stage of this DR. 
 
\begin{figure} [ht]
  \begin{center}
  \begin{tabular}{c} 
  \includegraphics[height=8cm]{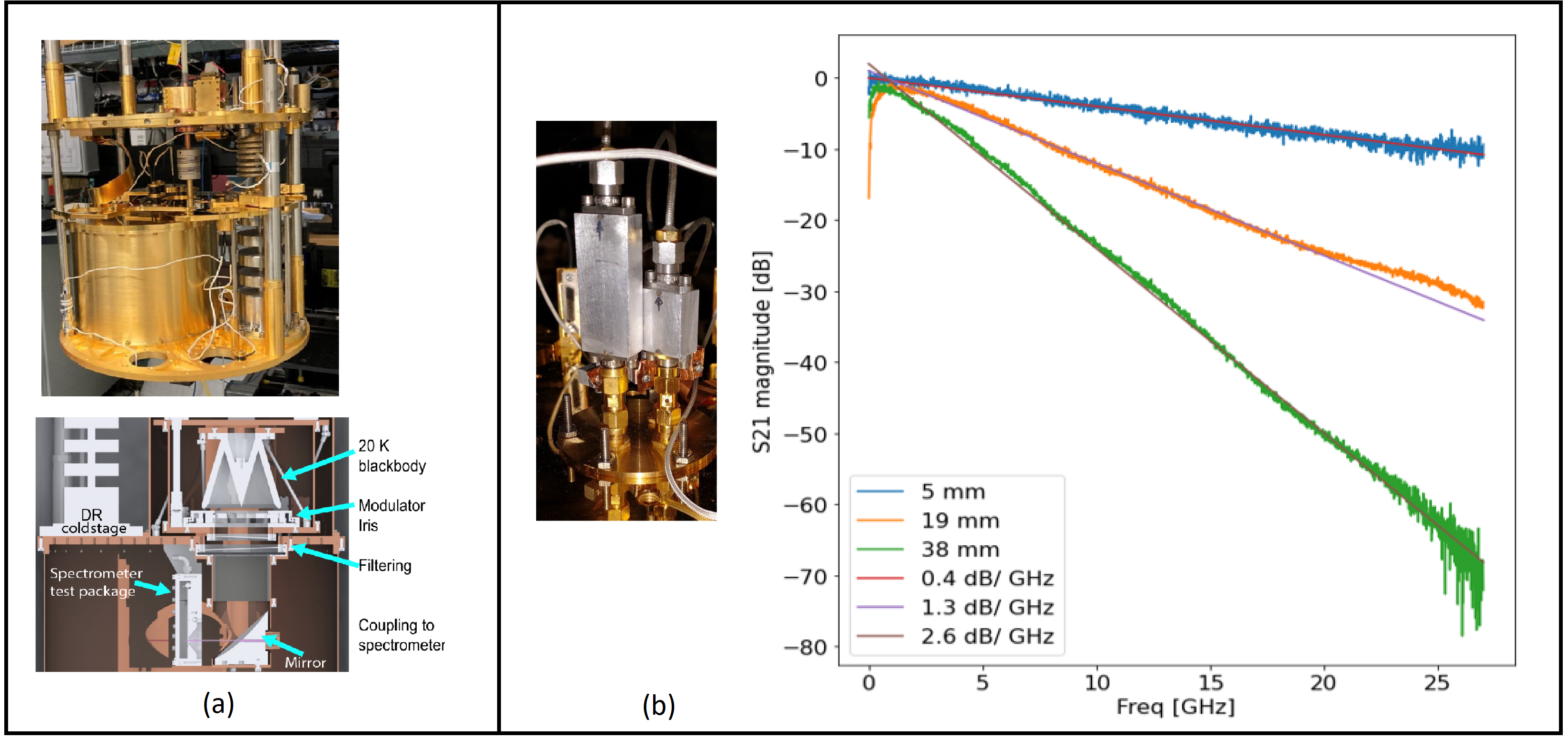}
  \end{tabular}
  \end{center}
  \caption[example] 
  { \label{fig:DR_testbed1} 
 a) Blackbody source mounted in the DR (top) and its components (bottom), b) thermal blocking powder filters and their transmission measured at room temperature.}
  \end{figure}

 RF lines route via 50\,K, 3\,K, and ${\sim}800$\,mK stages to this lowest stage. Two RF channels are installed, operating at frequency bands of $1-4$\, GHz and $4-8$\, GHz, as shown in Figure \ref{fig:RF_diagram}.  The two channels have some minor differences, as the $4-8$\, GHz channel employs a Low Noise Factory amplifier \cite{switzer2021experiment} ,LNF-LNC4-8C (a 42 dB gain, a cryogenic Inp HEMT amplifier), while the $1-4$\, GHz channel employs a Si-Ge amplifier, CITLF SN:45 ( a 46dB gain, cryogenic low noise amplifier) \cite{weinreb2007design,weinreb2009matched}. Both are mounted on the 3\,K stage. Other differences are the length (5.12, 19.1, and 38.9\,mm) of thermal blocking powder filters inserted at the coldest stage. As illustrated in Figure~\ref{fig:RF_diagram}, four powder filters are utilized for achieving an ultra-low background. These impedance-matched absorptive blocking filters block thermal radiation carried down the microwave lines using a lossy dielectric\cite{wollack2014impedance}. Figure \ref{fig:DR_testbed1}-b shows the transmission measured at room temperature (300\,K), which match design expectations\cite{wollack2014impedance}. The cables from 300\,K to 3\,K stage are beryllium-copper, and from 3\,K to the sample stage, NbTi cables are employed. An external cylindrical mu-metal shield is installed during cooldown around the full length of the cryostat to shield the superconducting detectors from Earth's field.
 
 The testbed includes options for readout with homodyne, VNA or DAC/ADC in combination with Open Architecture Computing Hardware (ROACH-2) system with a corresponding IF board interface. Through these setups, the feedline transmission S-parameter, $S_{21}$, can be probed or measured as a function of RF frequency across all of the resonators in the KID array to determine the KID resonance frequencies and quality factors as a function of temperature and readout tone power. The current laboratory ROACH-2 multiplexing readout system can be used to efficiently characterize the optical response of all the spectrometer KID channels simultaneously. 
 
 Two new optical facilities have recently been developed for this testbed, including an installation for coupling to a photomixer swept-source external to the cryostat and an internal low background beam-filling blackbody source. The internal blackbody source (depicted in Figure \ref{fig:DR_testbed1}) consists of a 1-20\,K blackbody, a modulating iris, and optical coupling to the spectrometer package through an off-axis parabolic mirror (OAP)\cite{Connorsposter}. The NEP and efficiency of future $\mu$-Spec's will be characterized using this low-background source, mounted internally to the cryostat. The blackbody simulates balloon and space-flight loading conditions, allowing determination of the noise performance of the $\mu$-Spec as a function of optical power and extraction of the $\mu$-Spec's optical efficiency. This blackbody has begun commissioning in the lab's DR test facility, and future publications will describe its design and performance. In the following sections, we focus only on the description of the design and functionality of the second new optical testbed, an external photomixer source configuration. This photomixer provides a swept line source that allows us to characterize the spectrometer's spectral response, both in-band and out-of-band.

\section{External Optical Testbed}
\label{sec:externalOpticalTestbed}
\subsection{Photomixer Source}
\begin{figure} [ht]
  \begin{center}
  \begin{tabular}{c} 
  \includegraphics[height=11.5cm]{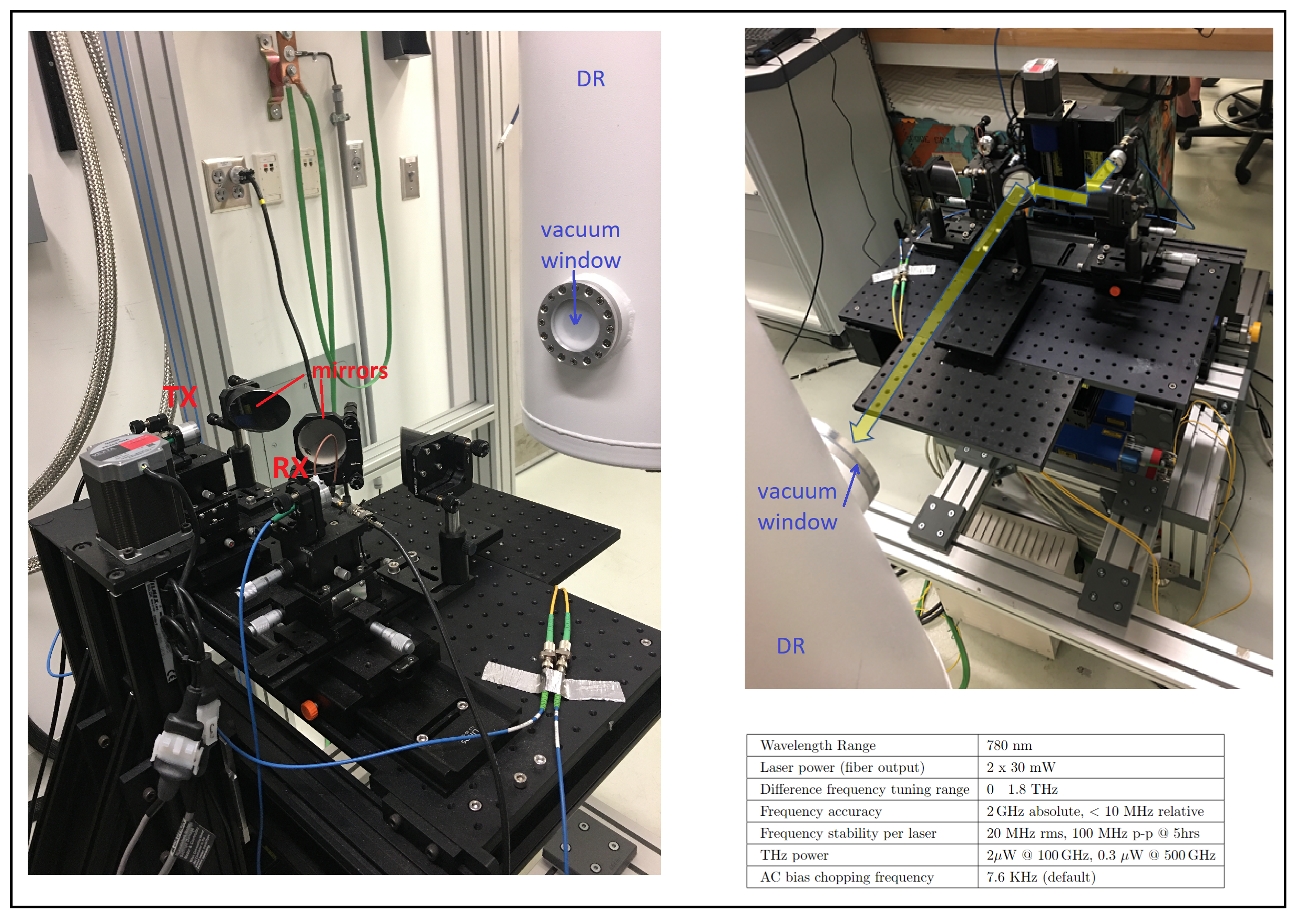}
  \end{tabular}
  \end{center}
  \caption[example] 
  { \label{fig:photom1} 
 The Toptica photomixer shown from two view angles, including the transmitter (TX) and receiver (RX), mirrors, the DR and the vacuum window mounted on it, and the moving stage.}
  \end{figure}

The external source used for this facility is a Toptica cw Terahertz GaAs photomixer \cite{Toptica_photom}. In this photomixer, a continuous-wave THz signal is generated by optical heterodyning in high-bandwidth photoconductors. The photomixer uses two infrared laser beams with a small frequency difference and mixes them into a GHz/THz wave. The outputs of the lasers irradiate the photomixer, which is made of metal-semiconductor-metal and creates charge carriers that, in turn, induce a voltage across two metal electrodes and generate a photocurrent that oscillates at a beat frequency equal to the difference between the two laser's frequencies. The antenna surrounding the photomixer translates this photocurrent into an emitted GHz or THz wave. In standard configuration, the Toptica photomixer is configured to be used with a room temperature receiver for transmission measurements. For operation in our testbed as a GHz/THz source, a 90-degree mirror is used to divert the emitted beam and transmit it via a window into our cryostat. The table in Fig.~\ref{fig:photom1} lists the main performance specifications which are utilized in this experimental setup \cite{Toptica_photom}. The photomixer source can be swept from optical wavelengths of 3\,mm to 166\,\um, which is done by controlling the respective temperatures of the two lasers.  Fig. \ref{fig:photom1} depicts this Toptica photomixer from two different views showing the transmitter (TX), receiver (RX), mirrors, and cryostat window. The photomixer is mounted on a custom three-dimensional programmable moving stage that allows for precise alignment well below the mm-scale. 

\begin{figure}[h]
   \begin{center}
      \includegraphics[height=7cm]{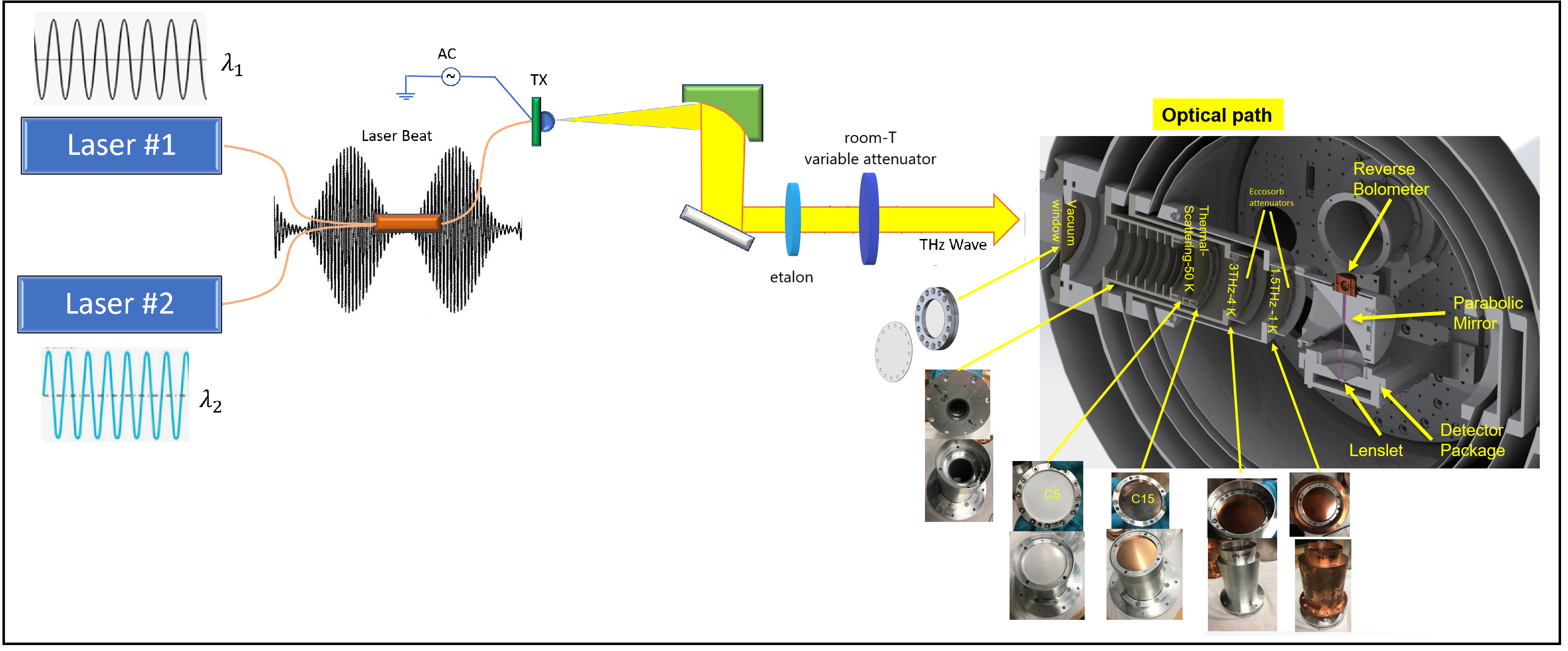}
   \end{center}
   \caption[example] 
   {\label{fig:photom_setup}The photomixer optical layout.}
\end{figure} 

\subsection{Optical Path}
The photomixer swept source couples to the spectrometers through optics into the low-background dilution refrigerator microwave testbed. Figure \ref{fig:photom_setup} depicts the components in this optical path, including the Toptica GaAs photomixer system components which couples the far-infrared wave into the DR and $\mu$-Spec through the vacuum window and filtering stages.

\begin{figure} [ht]
  \begin{center}
  \begin{tabular}{c} 
  \includegraphics[height=11.5cm]{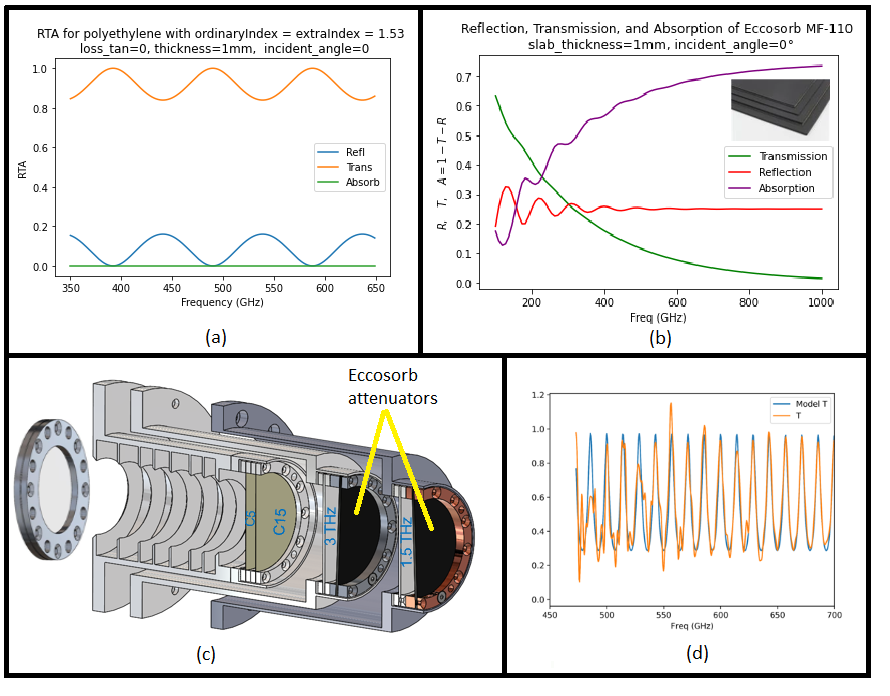}
  \end{tabular}
  \end{center}
  \caption[example] 
  { \label{fig:optical_layout1} 
 a) The vacuum window and its reflection, transmission, and absorption diagrams, b) Eccosorb attenuator and their transmission, reflection, and absorption, and c) The layout of the vacuum window and Eccosorb attenuators, and d) measured transmission versus optical frequency, with a model fit, of the Tydex etalon calibrator that will be used to calibrate the photomixer source absolute frequency.}
  \end{figure}

The DR vacuum window is an ultra-high molecular weight polyethylene (UHMWPE) with 1 mm thickness. The reflection, transmission, and absorption of the vacuum window with an ordinary index of 1.53, loss tangent of zero, and incident angle of zero are illustrated in Figure \ref{fig:optical_layout1}-a.

A baffle and series of thermal blocking filters\cite{pisano2015metal} are mounted inside the cryostat at the window to control thermal radiation from the warm room. The filters are provided by Cardiff University and include a 1.5\,THz cutoff filter at the 1\,K stage (the innermost stage), a 3\,THz cutoff filter at the 3\,K stage, and  dual-element thermal and scattering filters at the 50\,K stage. This baseline baffle and filter stack successfully limits the total thermal radiation incident on the DR mixing chamber stage to less than 40\,$\mu {\rm W}$, and the DR base temperature remains below 50\,mK when it is installed.

In addition to these thermal blocking filters, for tests with EXCLAIM spectrometers, custom cold Eccsorb attenuators \cite{Eccosorb-MF-Larid} are also installed. For this purpose, We have designed these attenuators made from 1\,mm-thick Eccosorb sheet, which provide 18-20\,dB of absorption. In Figure \ref{fig:optical_layout1}-b, the reflection, transmission, and absorption of Eccosorb MF-110 with a slab thickness of 1\,mm and incident angle of zero are shown. The following design equations \cite{stratton2007electromagnetic} were used to calculate these transmission parameters:\\
\begin{equation}
\label{eq:fov}
\phi = \frac{2\pi d}{\lambda_0} \sqrt{K^*.K_m^*-sin^2\theta}\, ,
\,\,\,\,\, r=\frac{\sqrt{K_m^*}-\sqrt{K^*}}{\sqrt{K_m^*}+\sqrt{K^*}} 
\end{equation}

\begin{equation}
\label{eq:fov}
T=\left|\frac{(1-r^2)e^{-j\phi}}{1-r^2 e^{-2j\phi}}\right|^2\, , \,\,\, R=\left|\frac{-r(1-e^{-2j\phi})}{1-r^2 e^{-2j\phi}}\right|^2\, , \,\,\, A= 1-T -R
\end{equation}

where $K^*$ is the complex dielectric constant, $K_m^*$ is the complex relative magnetic permeability, $\theta$ is the incident angle, $\lambda_0$ is the free-space wavelength, $\phi$ is the interface electrical thickness, $r_{\parallel}$ and $r_{\bot}$ are the interface's parallel and perpendicular voltage reflection coefficients, and $R$, $T$, and $A$ are the power reflection, transmission, and absorption coefficients. 

Additional variable attenuation is also provided at room temperature by Tydex THz attenuators, ATS-5-50.8 set consisting of five wheels of filters \cite{Tydex_var_Att}, four of them containing metalized wedged silicon wafers with different attenuation levels. The different combinations of these filters of -20 dB, -15.2 dB, -10 dB, and -5.23 dB can provide -5.23 dB to a minimum value of -50.5 dB of transmission. These attenuation levels as reported by Tydex on a range of wavelength from 10 to 1300 $\mu$m \cite{Tydex_var_Att}. 


A silicon etalon is introduced at room temperature after the photomixer mirror to calibrate the photomixer to more precisely determine the absolute frequency response of the $\mu$-Spec channels. The etalon TEFP-HRFZ-Si-D25.4-T3 is a HRFZ Si Fabry-Perot Etalon with a diameter of 7602\,mm and thickness of 3.0\,mm \cite{Tydex_Etalon}. Calibration of this etalon was performed via an FTS measurement at Goddard and is shown in Figure \ref{fig:optical_layout1}-d.\par

\subsection{Quasi-optical Coupling Design}

\begin{figure} [h]
   \begin{center}
   \begin{tabular}{c} 
   \includegraphics[height=7cm]{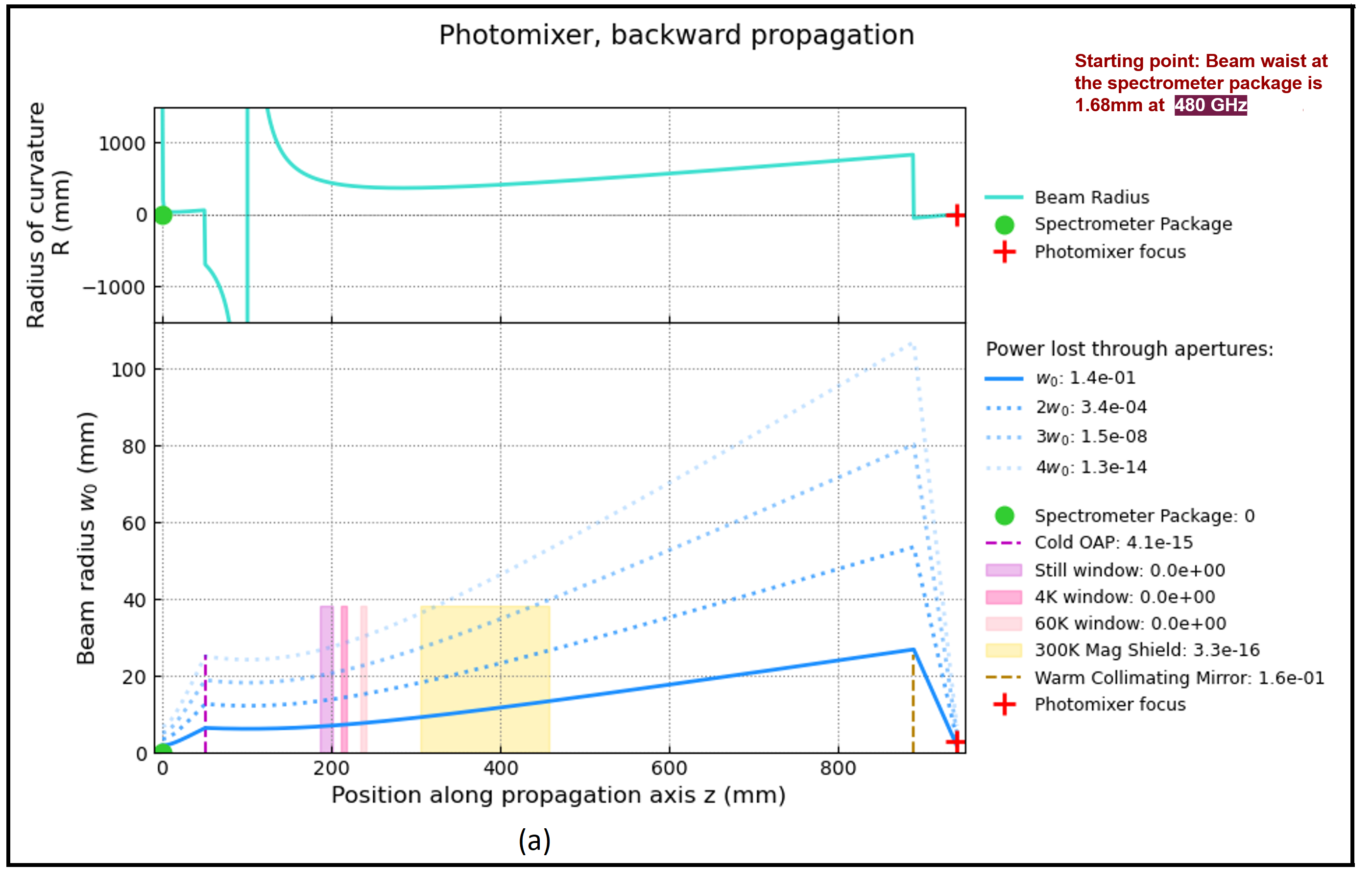}
   \end{tabular}
   \end{center}
   \caption[example] 
   { \label{fig:photom_Alyssa} 
Gaussian beam models of the DR lab test setup for $\mu$-Spec optical testing. }
   \end{figure} 
   
 To ensure efficient optical coupling between the photomixer and detector package, the optical performance of the system was modeled using a quasi-optical Gaussian beam approach. Figure~\ref{fig:photom_Alyssa} shows the Gaussian beam analysis of the photomixer setup, including power levels lost as the Gaussian beam propagates through the finite aperture and window sizes. Outside the cryostat, an off-axis parabolic (OAP) mirror collimates the light as the photomixer beam enters the cryostat through a series of windows. A second OAP on the cold stage refocuses the beam onto the spectrometer. In the Gaussian optics model, positions and aperture sizes of the optical elements are considered in order along the propagation axis. Our modeling demonstrates that negligible beam power is lost due to the Gaussian beam propagating through finite-diameter windows and apertures. The Gaussian beam analysis is used mainly to confirm that we get sufficiently good optical coupling from the spectrometer to the photomixer. Any light that is not well-coupled could produce negative impacts such as loading or reflections. 
 
\section{Test Plan With External Source}
\label{sec:TestplanR64}

The photomixer provides a swept line source allowing characterization of the spectrometer resolution and each spectrometer channel's absolute frequency response. The absolute frequency response can be calibrated within the requirement of $\pm 1$\,GHz needed for EXCLAIM science by completing an optical frequency scan with the etalon inserted. The photomixer also provides a method to electrically chop the optical signal, allowing us to measure (within error bars set by the optical load uncertainty) the KID time constants and to perform a lock-in measurement on the modulated signal. The photomixer also allows us to quantify contributions from stray light and Lorentzian or physical crosstalk by completing broadband frequency scans and monitoring the response of specific channels to out-of-band light.

\section{Results}
\label{sec:Results}

\begin{figure} [h]
   \begin{center}
   \begin{tabular}{c} 
   \includegraphics[height=5cm]{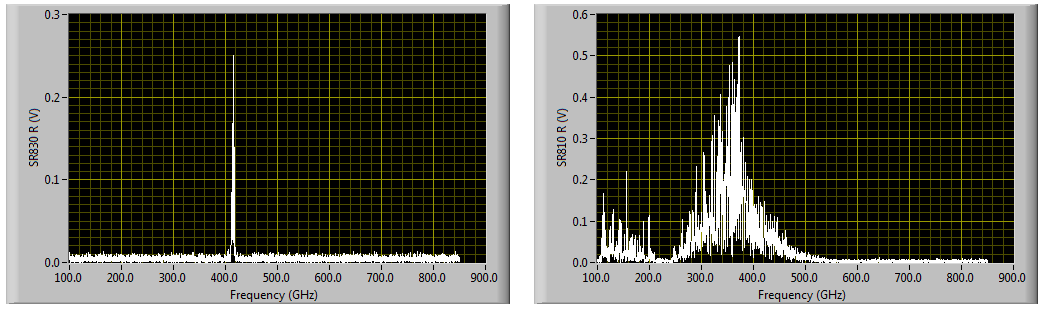}
   \end{tabular}
   \end{center}
   \caption[example] 
   { \label{fig:broadband_measurement} 
R64 resolution demonstration measured by Toptica photomixer a) $\mu$-Spec's broadband response of the resonator channel 1, b) the broadband response of the reference resonator.}
   \end{figure}

The optical components described in Sect.~\ref{sec:externalOpticalTestbed} have been installed in the DR testbed. And an initial validation of the system was performed on a R=64 prototype spectrometer device. Optical frequency scans of spectrometer channels were performed simultaneously with a reference KID channel to normalize the response. Figure \ref{fig:broadband_measurement}-a and -b depicts the measured response of the 1st channel of the spectrometer and the reference resonator, respectively.

\begin{figure} [h]
   \begin{center}
   \begin{tabular}{c} 
   \includegraphics[height=5cm]{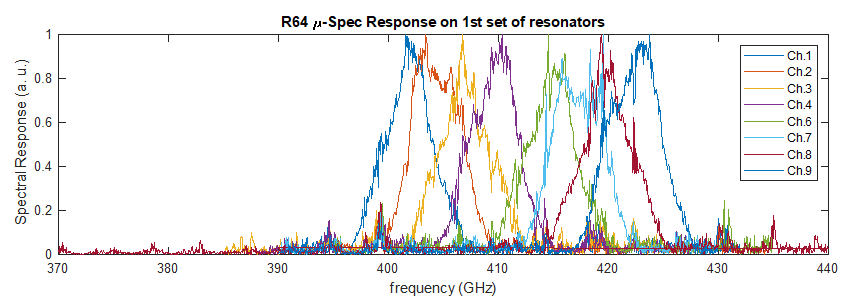}
   \end{tabular}
   \end{center}
   \caption[example] 
   { \label{fig:R64_response_measurement} 
R64 frequency resolution response measured by photomixer, two lock-in-amplifiers to measure the reference channel and the spectrometer channels at the same time. The diagram is the normalized response of the spectrometer's resonators versus the reference resonator response.}
   \end{figure}

Note that in these measurements of the prototype R64 spectrometer, given its lower sensitivity, we do not use the attenuators or variable attenuators. The measurements performed allow us to characterize the spectral and absolute frequency response. Figure \ref{fig:R64_response_measurement} illustrates the absolute frequency and resolution of the $\mu$-Spec response which is incorporated by Toptica photomixer. The absolute optical frequency response of the spectrometer channels are shifted down by 9 GHz due to variation in kinetic inductance from design assumptions, while the resolution and line profile are as designed. For this measurement a homodyne readout system was used where two signal sources were used to generate the required LO signals with the frequency adjusted to that of the reference resonator and the desired spectrometer channel at a defined power level. The resulting I/Q signals at the outputs of two demodulators after passing through a pre-amplifier and filtering system are sent to two lock-in-amplifiers to readout the response. The 7kHz modulated bias signal from the photomixer THz wave is sent to the lock-in-amplifiers as the reference signals. The outputs of the lock-in-amplifiers are then recorded  using a Labview software program as shown in Figure \ref{fig:broadband_measurement}.

\section{Conclusions}


We have described the development of a new optical testbed for the characterization of $\mu$-Spec spectrometers. An overview of the DR testbed and details of an external optical source (photomixer) test setup were presented. Employing the photomixer test setup, the absolute frequency response and spectral resolution can be determined. Moreover, an existing blackbody test setup mounted in the DR can be used for efficiency and NEP measurements.  Initial validation of this external optical coupling setup was demonstrated with a R=64 $\mu$-Spec. Further validation (including validation using the ROACH2 simultaneous readout) is currently in progress, and is anticipated leading up its use for characterization of the R=512 EXCLAIM $\mu$-Specs. The EXCLAIM spectrometer testplan will include characterization using both dark, blackbody source, and photomixer measurements. 

\acknowledgments 
Financial support received from the NASA ROSES/APRA Detector Development Program (R=512 Design Development) 2016-2019 ROSES/APRA Balloon Program (EXCLAIM) 2019 -current and the Oak Ridge Associated Universities (ORAU) fellowship are gratefully acknowledged by the authors.

\bibliography{report} 

\begin{thebibliography}{10}

\bibitem{switzer2021experiment}
Switzer, E.~R., Barrentine, E.~M., Cataldo, G., Essinger-Hileman, T., Ade,
  P.~A., Anderson, C.~J., Barlis, A., Beeman, J., Bellis, N., Bolatto, A.~D.,
  et~al., ``Experiment for cryogenic large-aperture intensity mapping:
  instrument design,'' {\em Journal of Astronomical Telescopes, Instruments,
  and Systems}~{\bf 7}(4),  044004 (2021).

\bibitem{endo2012development}
Endo, A., Baselmans, J., van~der Werf, P., Knoors, B., Javadzadeh, S., Yates,
  S., Thoen, D., Ferrari, L., Baryshev, A., Lankwarden, Y., et~al.,
  ``Development of deshima: a redshift machine based on a superconducting
  on-chip filterbank,'' in [{\em Millimeter, Submillimeter, and Far-Infrared
  Detectors and Instrumentation for Astronomy VI}{\nolinebreak\hspace{0.1em}]},
    {\bf 8452},  253--267, SPIE (2012).

\bibitem{karkare2020full}
Karkare, K., Barry, P., Bradford, C., Chapman, S., Doyle, S., Glenn, J.,
  Gordon, S., Hailey-Dunsheath, S., Janssen, R., Kov{\'a}cs, A., et~al.,
  ``Full-array noise performance of deployment-grade superspec mm-wave on-chip
  spectrometers,'' {\em Journal of Low Temperature Physics}~{\bf 199}(3),
  849--857 (2020).

\bibitem{robson2022simulation}
Robson, G., Anderson, A.~J., Barry, P.~S., Doyle, S., and Karkare, K.~S., ``The
  simulation and design of an on-chip superconducting millimetre filter-bank
  spectrometer,'' {\em Journal of Low Temperature Physics} ,  1--9 (2022).

\bibitem{thomas2014cambridge}
Thomas, C.~N., Withington, S., Maiolino, R., Goldie, D.~J., Acedo, E., Wagg,
  J., Blundell, R., Paine, S., and Zeng, L., ``The cambridge emission line
  surveyor (camels),'' {\em arXiv preprint arXiv:1401.4395}  (2014).

\bibitem{basu2020superconducting}
Basu~Thakur, R., Klimovich, N., Day, P., Shirokoff, E., Mauskopf, P.,
  Faramarzi, F., and Barry, P., ``Superconducting on-chip fourier transform
  spectrometer,'' {\em Journal of Low Temperature Physics}~{\bf 200}(5),
  342--352 (2020).

\bibitem{cataldo2019second}
Cataldo, G., Barrentine, E.~M., Bulcha, B.~T., Ehsan, N., Hess, L.~A.,
  Noroozian, O., Stevenson, T.~R., Wollack, E.~J., Moseley, S.~H., and Switzer,
  E.~R., ``Second-generation micro-spec: A compact spectrometer for
  far-infrared and submillimeter space missions,'' {\em Acta Astronautica}~{\bf
  162},  155--159 (2019).

\bibitem{cataldo2020overview}
Cataldo, G., Ade, P.~A., Anderson, C.~J., Barlis, A., Barrentine, E.~M.,
  Bellis, N.~G., Bolatto, A.~D., Breysse, P.~C., Bulcha, B.~T., Connors, J.~A.,
  et~al., ``Overview and status of exclaim, the experiment for cryogenic
  large-aperture intensity mapping,'' in [{\em Ground-based and Airborne
  Telescopes VIII}{\nolinebreak\hspace{0.1em}]},   {\bf 11445},  469--479, SPIE
  (2020).

\bibitem{mirzaei2020mu}
Mirzaei, M., Barrentine, E.~M., Bulcha, B.~T., Cataldo, G., Connors, J.~A.,
  Ehsan, N., Essinger-Hileman, T.~M., Hess, L.~A., Mugge-Durum, J.~W.,
  Noroozian, O., et~al., ``$\mu$-spec spectrometers for the exclaim
  instrument,'' in [{\em Millimeter, Submillimeter, and Far-Infrared Detectors
  and Instrumentation for Astronomy X}{\nolinebreak\hspace{0.1em}]},   {\bf
  11453},  128--139, SPIE (2020).

\bibitem{bulcha2018electromagnetic}
Bulcha, B., Cataldo, G., Stevenson, T., U-Yen, K., Moseley, S., and Wollack,
  E., ``Electromagnetic design of a magnetically coupled spatial power
  combiner,'' {\em Journal of Low Temperature Physics}~{\bf 193}(5),  777--785
  (2018).

\bibitem{baselmans2017kilo}
Baselmans, J., Bueno, J., Yates, S.~J., Yurduseven, O., Llombart, N., Karatsu,
  K., Baryshev, A., Ferrari, L., Endo, A., Thoen, D., et~al., ``A kilo-pixel
  imaging system for future space based far-infrared observatories using
  microwave kinetic inductance detectors,'' {\em Astronomy \&
  Astrophysics}~{\bf 601},  A89 (2017).

\bibitem{hailey2021kinetic}
Hailey-Dunsheath, S., Janssen, R.~M., Glenn, J., Bradford, C.~M., Perido, J.,
  Redford, J., and Zmuidzinas, J., ``Kinetic inductance detectors for the
  origins space telescope,'' {\em Journal of Astronomical Telescopes,
  Instruments, and Systems}~{\bf 7}(1),  011015 (2021).

\bibitem{mazin2009microwave}
Mazin, B.~A., ``Microwave kinetic inductance detectors: The first decade,'' in
  [{\em AIP Conference Proceedings}{\nolinebreak\hspace{0.1em}]},   {\bf
  1185}(1),  135--142, American Institute of Physics (2009).

\bibitem{zmuidzinas2012superconducting}
Zmuidzinas, J., ``Superconducting microresonators: Physics and applications,''
  {\em Annu. Rev. Condens. Matter Phys.}~{\bf 3}(1),  169--214 (2012).

\bibitem{bradley2021advancements}
Bradley, D.~C., Jamison-Hooks, T.~L., Staguhn, J.~G., Amatucci, E.~G.,
  Browning, T., DiPirro, M.~J., Leisawitz, D.~T., and Carter, R.~C., ``On the
  advancements of digital signal processing hardware and algorithms enabling
  the origins space telescope,'' {\em Journal of Astronomical Telescopes,
  Instruments, and Systems}~{\bf 7}(1),  011018 (2021).

\bibitem{meixner2019origins}
Meixner, M., Cooray, A., Leisawitz, D., Staguhn, J., Armus, L., Battersby, C.,
  Bauer, J., Bergin, E., Bradford, C., Ennico-Smith, K., et~al., ``Origins
  space telescope mission concept study report,'' {\em arXiv preprint
  arXiv:1912.06213}  (2019).

\bibitem{sinclair2020development}
Sinclair, A., Stephenson, R., Hoh, J., Gordon, S., and Mauskopf, P., ``On the
  development of a reconfigurable readout for superconducting arrays,'' in
  [{\em Millimeter, Submillimeter, and Far-Infrared Detectors and
  Instrumentation for Astronomy X}{\nolinebreak\hspace{0.1em}]},   {\bf 11453},
   243--251, SPIE (2020).

\bibitem{uSpec_Barrentine_2016}
{Barrentine}, E.~M., {Cataldo}, G., {Brown}, A.~D., {Ehsan}, N., {Noroozian},
  O., {Stevenson}, T.~R., {U-Yen}, K., {Wollack}, E.~J., and {Moseley}, S.~H.,
  ``{Design and performance of a high resolution {\ensuremath{\mu}}-spec: an
  integrated sub-millimeter spectrometer},'' in [{\em Millimeter,
  Submillimeter, and Far-Infrared Detectors and Instrumentation for Astronomy
  VIII}{\nolinebreak\hspace{0.1em}]},  {Holland}, W.~S. and {Zmuidzinas}, J.,
  eds., {\em Society of Photo-Optical Instrumentation Engineers (SPIE)
  Conference Series} {\bf 9914},  99143O (July 2016).

\bibitem{noroozian2015mu}
Noroozian, O., Barrentine, E., Brown, A., Cataldo, G., Ehsan, N., Hsieh, W.-T.,
  Stevenson, T., U-yen, K., Wollack, E., and Moseley, S.~H., ``$\mu$-spec: An
  efficient compact integrated spectrometer for submillimeter astrophysics,''
  in [{\em 26th International Symposium On Space Terahertz
  Technology}{\nolinebreak\hspace{0.1em}]},  (2015).

\bibitem{barrentine2015overview}
Barrentine, E.~M., Noroozian, O., Brown, A.~D., Cataldo, G., Ehsan, N., Hsieh,
  W.-T., Stevenson, T.~R., U-Yen, K., Wollack, E.~J., and Moseley, S.~H.,
  ``Overview of the design, fabrication and performance requirements of
  micro-spec, an integrated submillimeter spectrometer,'' in [{\em
  International Workshop on Low Temperature
  Detectors}{\nolinebreak\hspace{0.1em}]},  (GSFC-E-DAA-TN25496) (2015).

\bibitem{cataldo2014micro}
Cataldo, G., Hsieh, W.-T., Huang, W.-C., Moseley, S.~H., Stevenson, T.~R., and
  Wollack, E.~J., ``Micro-spec: an ultracompact, high-sensitivity spectrometer
  for far-infrared and submillimeter astronomy,'' {\em Applied optics}~{\bf
  53}(6),  1094--1102 (2014).

\bibitem{EXCLAIM_Switzer_JATIS_2021}
{Switzer}, E.~R., {Barrentine}, E.~M., {Cataldo}, G., {Essinger-Hileman}, T.,
  {Ade}, P. A.~R., {Anderson}, C.~J., {Barlis}, A., {Beeman}, J., {Bellis}, N.,
  {Bolatto}, A.~D., {Breysse}, P.~C., {Bulcha}, B.~T., {Chevres-Fernanadez},
  L.-R., {Cho}, C., {Connors}, J.~A., {Ehsan}, N., {Glenn}, J., {Golec}, J.,
  {Hays-Wehle}, J.~P., {Hess}, L.~A., {Jahromi}, A.~E., {Jenkins}, T.,
  {Kimball}, M.~O., {Kogut}, A.~J., {Lowe}, L.~N., {Mauskopf}, P., {McMahon},
  J., {Mirzaei}, M., {Moseley}, H., {Mugge-Durum}, J., {Noroozian}, O.,
  {Oxholm}, T.~M., {Parekh}, T., {Pen}, U.-L., {Pullen}, A.~R., {Rahmani}, M.,
  {Ramirez}, M.~M., {Roselli}, F., {Shire}, K., {Siebert}, G., {Sinclair},
  A.~K., {Somerville}, R.~S., {Stephenson}, R., {Stevenson}, T.~R., {Timbie},
  P., {Termini}, J., {Trenkamp}, J., {Tucker}, C., {Visbal}, E., {Volpert},
  C.~G., {Wollack}, E.~J., {Yang}, S., and {Yung}, L.~Y.~A., ``{Experiment for
  cryogenic large-aperture intensity mapping: instrument design},'' {\em
  Journal of Astronomical Telescopes, Instruments, and Systems}~{\bf 7},
  044004 (Oct. 2021).

\bibitem{weinreb2007design}
Weinreb, S., Bardin, J.~C., and Mani, H., ``Design of cryogenic sige low-noise
  amplifiers,'' {\em IEEE transactions on microwave theory and techniques}~{\bf
  55}(11),  2306--2312 (2007).

\bibitem{weinreb2009matched}
Weinreb, S., Bardin, J., Mani, H., and Jones, G., ``Matched wideband low-noise
  amplifiers for radio astronomy,'' {\em Review of Scientific Instruments}~{\bf
  80}(4),  044702 (2009).

\bibitem{wollack2014impedance}
Wollack, E., Chuss, D., Rostem, K., and U-Yen, K., ``Impedance matched
  absorptive thermal blocking filters,'' {\em Review of Scientific
  Instruments}~{\bf 85}(3),  034702 (2014).

\bibitem{Connorsposter}
Connors~et al., J., ``Towards photon counting kinetic inductance detectors for
  far-infrared spectroscopy.'' Poster presentation at 18th International
  Workshop on Low Temperature Detectors (July 2019).

\bibitem{Toptica_photom}
Toptica, ``{cw THz photomixer}.'' Toptica website,
  \url{https://www.toptica.com/products/terahertz-systems/frequency-domain/gaas-and-ingaas-photomixers/}.
\newblock (Accessed: 9 July 2022).

\bibitem{pisano2015metal}
Pisano, G., Tucker, C., Ade, P.~R., Moseley, P., and Ng, M.~W., ``Metal mesh
  based metamaterials for millimetre wave and thz astronomy applications,'' in
  [{\em 2015 8th UK, Europe, China Millimeter Waves and THz Technology Workshop
  (UCMMT)}{\nolinebreak\hspace{0.1em}]},   1--4, IEEE (2015).

\bibitem{Eccosorb-MF-Larid}
Larid, ``{Eccosorb Attenuators}.'' Larid website,
  \url{https://www.laird.com/sites/default/files/2021-07/RFP-DS-MF%20061721.pdf}.
\newblock (Accessed: 12 July 2022).

\bibitem{stratton2007electromagnetic}
Stratton, J.~A.,  [{\em Electromagnetic theory}{\nolinebreak\hspace{0.1em}]},
  vol.~33, John Wiley \& Sons (2007).

\bibitem{Tydex_var_Att}
Tydex, C., ``{cw Terahertz Spectroscopy Systems, Manual}.'' Tydex,
  \url{http://www.tydexoptics.com/pdf/THz_Attenuators.pdf}.
\newblock (Accessed: 5 July 2022).

\bibitem{Tydex_Etalon}
Tydex, C., ``{cw Terahertz Spectroscopy Systems, Manual}.'' Tydex, \url{
  http://www.tydexoptics.com/pdf/THz_F-P_Etalon.pdf}.
\newblock (Accessed: 5 July 2022).

\end{thebibliography}
\bibliographystyle{spiebib} 

\end{document}